\documentclass[twocolumn,amsmath,amssymb,superscriptaddress,prb]{revtex4-1}

\usepackage[dvipdfmx]{graphicx}
\usepackage{graphicx}% Include figure files
\usepackage{dcolumn}% Align table columns on decimal point
\usepackage{bm}% bold math
\usepackage{color}
\def\Vec#1{\bm{#1}}

\begin{document}

%\preprint{APS/123-QED}

\title{
Intrinsic vortex pinning in superconducting quasicrystals
}% Force line breaks with \\

\author{Yuki Nagai}
\email{nagai.yuki@jaea.go.jp }
\affiliation{CCSE, Japan  Atomic Energy Agency, 178-4-4, Wakashiba, Kashiwa, Chiba, 277-0871, Japan}
\affiliation{
Mathematical Science Team, RIKEN Center for Advanced Intelligence Project (AIP), 1-4-1 Nihonbashi, Chuo-ku, Tokyo 103-0027, Japan
}

\date{\today}% It is always \today, today,

\begin{abstract}
We numerically show that a vortex pinning occurs in superconducting quasicrystal without impurities and defects.
This vortex pinning is intrinsic since the superconducting order parameter in quasicrystal is always inhomogeneous due to the lack of the translational symmetry. 
We propose that experiments influenced by vortex pinning effects can detect the atomic-scale inhomogeneous superconducting order parameter in quasicrystals. 
We develop a numerical method to solve the Bogoliubov-de Gennes equations and gap equations in large systems, which is based on the localized-Krylov subspace and a sparse modeling technique. 
Two two-dimensional quasicrystals, the Penrose and Amman-Beenker tiling, are considered.

\end{abstract}

\maketitle

\section{Introduction}
Quasicrystal has no translational symmetry. 
Recently, Kamiya {\it et al.} found that a superconducting phase appears in Al–Zn–Mg quasicrystalline alloys\cite{Kamiya}. 
In superconducting quasicrystals, the Cooper pairs with momentum $k$ and $-k$, which appeared in the conventional picture of the superconductivity, can not be applied, since momentum is not a good quantum number.
The lack of the translational symmetry induces an intrinsic atomic-scale inhomogeneous superconducting order parameter in superconducting quasicrystals \cite{SakaiPRB,SakaiPRR}. 

The detection of the inhomogeneous phenomena in superconducting quasicrystals in experiments is one of the most important evidence of the existence of exotic superconductivity in quasicrystals. 
However, there is no experiment that can directly detect a superconducting order parameter in real space. 
Some theoretical papers found that the local density of states (LDOS), which can be observed by the scanning tunneling microscopy/spectroscopy (STM/STS), has only weak spatial dependence of the spectral gap \cite{SakaiPRB,NagaiLK}, even if the superconducting order parameter fluctuates in atomic scales. 

A vortex in superconductors can be directly observed by STS/STS or SQUID experiments. 
A vortex has been used to study superconducting properties, since bound states around a vortex reflect the information of the superconducting order parameter and the electronic structure. 
For example, a vortex in NbSe$_2$ has a six-fold star-shaped LDOS originating from a six-fold anisotropic superconducting pairing symmetry\cite{Hess,HayashiPRL}. 
Recently, Fermi surface anisotropy in conventional superconductor lanthanum can be detected by the STM/STM\cite{HowonKim}. 

With the use of a vortex in superconducting quasicrystals, we propose that the inhomogeneous superconducting state in quasicrystal can be detected.
We focus on the fact that the energy level of the minimum vortex bound states depends on how the  superconducting order parameter is suppressed around a vortex since a region where the order parameter becomes small around a vortex can be regarded as a region inside a quantum well for quasiparticles.  
In general, the energy level becomes small if the size of the suppressed region is large ({\it i.e.} inside a large quantum well).
In addition, the size of the suppressed region depends on the amplitude of the superconducting order parameter. 
Therefore, since the amplitude of the order parameter changes in superconducting quasicrystal in real space, the energy level of the vortex bound states should depend on the position of the vortex. 

In this paper, we show an intrinsic vortex pinning due to a strong inhomogeneous superconducting order parameter in superconducting quasicrystal, which can be detected experimentally. 
A vortex moves to find a position where the free energy becomes minimum. 
In other words, vortices are pinned even if there is no impurity or defect. 
To make this statement general, we consider two kinds of superconducting quasicrystals, superconducting tight-binding models with Penrose and Ammann-Beenker (AB) tiling. 
A new numerical technique is proposed for solving the Bogoliubov-de Gennes equations in a large inhomogeneous tight-binding model, which is based on the localized-Krylov subspace and sparse modeling techniques\cite{NagaiLK}.

This paper is organized as follows. 
In Sec.~\ref{sec:model}., we show the model and method that we consider. 
We show the theoretical model Hamiltonian of the two-dimensional tight-binding Bardeen-Cooper-Schrieffer (BCS) Hamiltonian on the Penrose and Ammann-Beenker tiling. We propose a new numerical approach for large-scale superconductors, localized-Krylov reduced shifted conjugate gradient method with sparse modeling (LK-RSCG with SpM), which is based on the localized-Krylov subspace and sparse modeling techniques.  We introduce the dual-grid method to construct models for quasicrystals, which is based on a projection from a high-dimensional lattice. With the use of this method, one can easily generate different patterns of tiling. 
In Sec.~\ref{sec:results}, we show the numerical results that the intrinsic vortex pinning occurs in superconducting quasicrystals. 
In Sec.~\ref{sec:discussions}, we discuss the mechanism of the intrinsic vortex pinning and its position. 
In Sec.~\ref{sec:summary}, the summary are given.

\section{Model and method}\label{sec:model}
\subsection{Hamiltonian}
We consider the tight-binding Bardeen-Cooper-Schrieffer (BCS) Hamiltonian on the Penrose and Ammann-Beenker tiling given as 
\begin{align}
{\cal H} &=  \sum_{ij,\sigma} (-t_{ij} - \mu \delta_{ij}) c_{i\sigma}^{\dagger} c_{j\sigma} + \sum_{i} \left[ \Delta_{i} c_{i \uparrow}^{\dagger} c_{i \downarrow}^{\dagger} + {\rm H.c.} \right],
\end{align}
where $c_{i\sigma}^{\dagger}$ creates the electron with spin $\sigma$ at site $i$ and $\mu$ denotes the chemical potential. 
$t_{ij}$ is the transfer integral which connects a bond on the tiling. 
We consider that the intensity of the hopping is uniform $t_{ij} = t$ on bonds. 
For simplicity, we consider on-site $s$-wave superconductivity. 
We use the unit system with $\hbar = k_{\rm B} = 1$. 
One can diagonalize ${\cal H}$ to solve the BdG equations expressed as 
\begin{align}
\sum_j  \hat{H}_{i,j}
\left(\begin{array}{c}
u_{\gamma}({\bm r}_j) \\
v_{\gamma}({\bm r}_j)
\end{array}\right)
&= 
E_{\gamma}
\left(\begin{array}{c}
u_{\gamma}({\bm r}_i) \\
v_{\gamma}({\bm r}_i)
\end{array}\right), \label{eq:bdg}
\end{align}
where the $2N \times 2N$ Hamiltonian matrix  $\hat{H}$ is defined as 
\begin{align}
    \hat{H}_{i,j} = \left(\begin{array}{cc}
[\hat{H}^{\rm N}]_{ij} & \Delta_{i} \delta_{ij} \\
\Delta_{i}^{\ast} \delta_{ij} & -[\hat{H}^{\rm N \ast}]_{ij} 
\end{array}\right).
\end{align}
Here $[\hat{H}^{\rm N}]_{ij} = - t_{ij} -\mu \delta_{ij}$ and 
%The size of the Hamiltonian matrix $\hat{H}$ is $2N$, where
$N$ is the number of the lattice sites. 
The $s$-wave superconducting order parameter is defined as 
\begin{align}
\Delta_i =U \langle c_{i\downarrow} c_{i \uparrow}\rangle, \label{eq:delta}
\end{align}
where $U$ is the onsite pairing interaction. 
In Eq.~(\ref{eq:delta}), we assume that the onsite pairing interaction does not depend on the lattice site. 
Although it seems more reasonable that the interaction have the lattice site dependence, the strong inhomogeneity of the superconducting order parameter has reported in both Penrose and Ammann-Beenker quasicrystals with site-independent pairing interactions\cite{Araujo,SakaiPRB}.
In the mean-field level, the physical properties of the systems are directly determined by the mean-fields, not the interactions. 
Phenomena induced by the site-dependent pairing interaction in the disordered Hubbard model are out of scope in this paper.

\subsection{Construction of quasicrystals: Dual-grid method}
We introduce the dual-grid method to construct models for quasicrystals\cite{Bruijn1981part1,Bruijn1981part2}. 
In this paper, we consider the Penrose quasicrystal and Ammann-Beenker quasicrystal, which are famous two-dimensional quasicrystals. 
There are several methods to construct models for quasicrystals. 
In last five years, the inflation-deflation method\cite{Levine} has been used to treat a large size cluster \cite{SakaiPRB,NagaiLK,SakaiPRR,Koga2020,Takemori2020,Mace2017,ChenPRL,Ghadimi,Koga2017}. 
In the inflation-deflation method, a quasicrystal is generated by iteratively applying the inflation-deflation rule. 
Usually, the quasicrystal generated by the inflation-deflation method has a high symmetric point at a center of a lattice. 
For example, in the Penrose quasicrystal, there is a ten-fold rotational symmetry around a center. 
However, in realistic quasicrystals, we can not assume that a rotational symmetry exists around a center of a system.

\begin{figure}[t]
    %%%%--- I comment out figure regions
    %\vspace{50mm}
    \begin{center}
         \begin{tabular}{p{ 0.5 \columnwidth} p{0.5 \columnwidth}}%  p{28mm}}
          (a) \resizebox{0.5 \columnwidth}{!}{\includegraphics{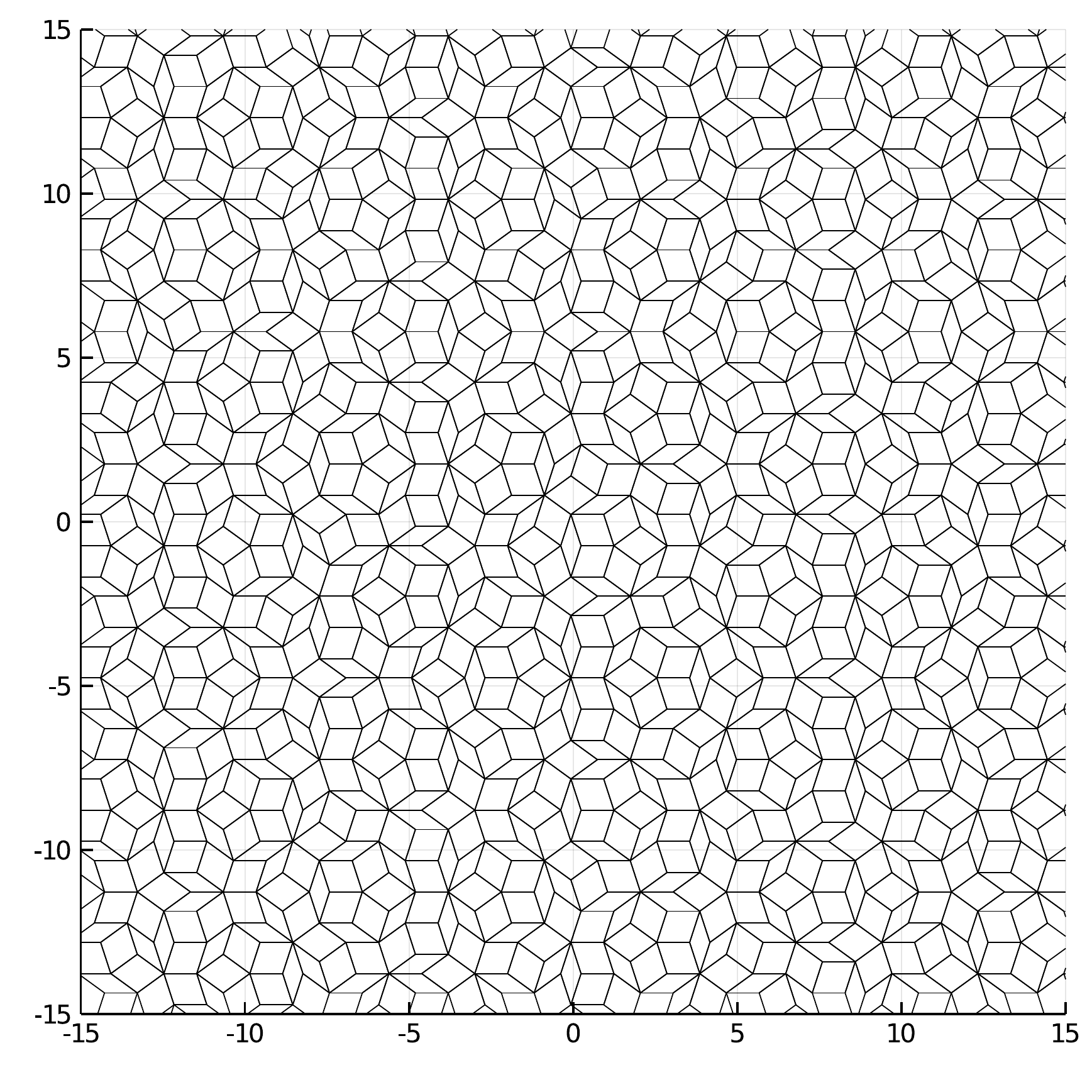}} 
          &
          (b) \resizebox{0.5 \columnwidth}{!}{\includegraphics{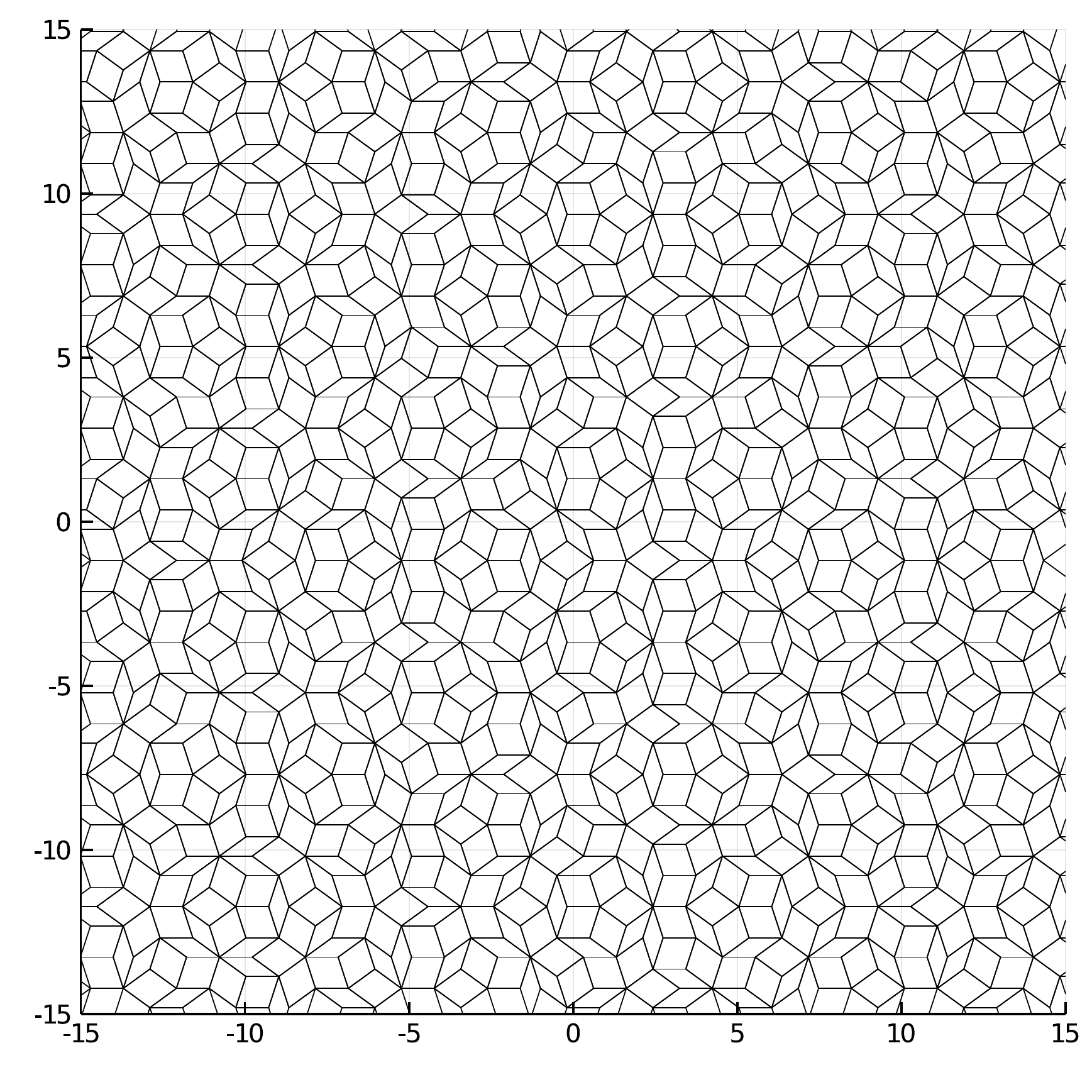}} 
        \end{tabular}
    \end{center}
    \caption{Penrose lattices obtained by the dual grid method with different $\vec{\gamma}^{\rm T}$. (a): $\vec{\gamma}^{\rm T}= (0.1,0.7,-0.98,0.43,-\sum_{\mu=0}^3\gamma_{\mu})$ and (b): $\vec{\gamma}^{\rm T}= (1/7,1/9,-3/4,-\sqrt{5}, -\sum_{\mu=0}^3\gamma_{\mu})$. 
    The total numbers of lattice sites are 60831 and 60863, respectively. 
    \label{fig:pen}
     }
    \end{figure}

\begin{figure}[t]
    %%%%--- I comment out figure regions
    %\vspace{50mm}
    \begin{center}
         \begin{tabular}{p{ 0.5 \columnwidth} p{0.5 \columnwidth}}%  p{28mm}}
          (a) \resizebox{0.5 \columnwidth}{!}{\includegraphics{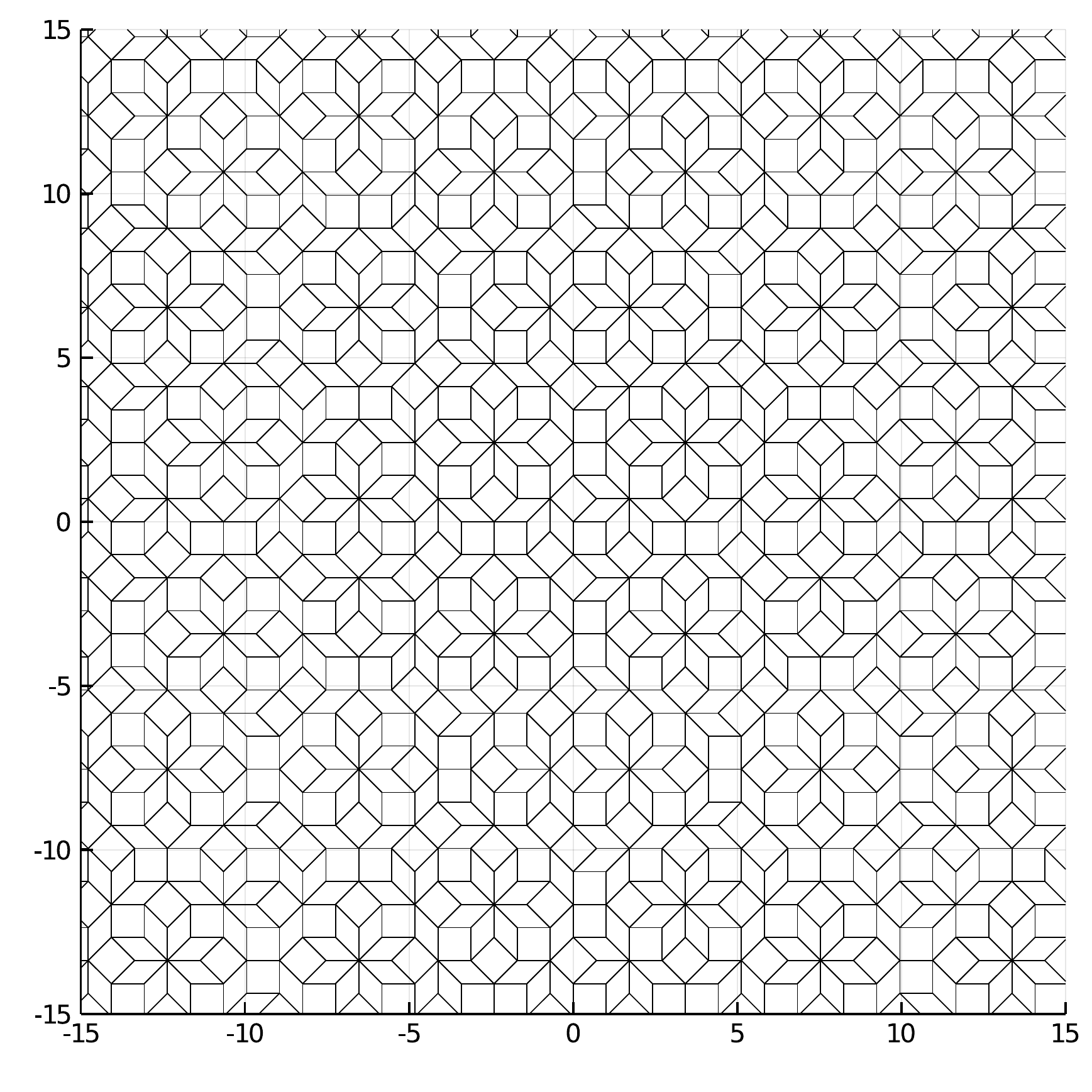}} 
          &
          (b) \resizebox{0.5 \columnwidth}{!}{\includegraphics{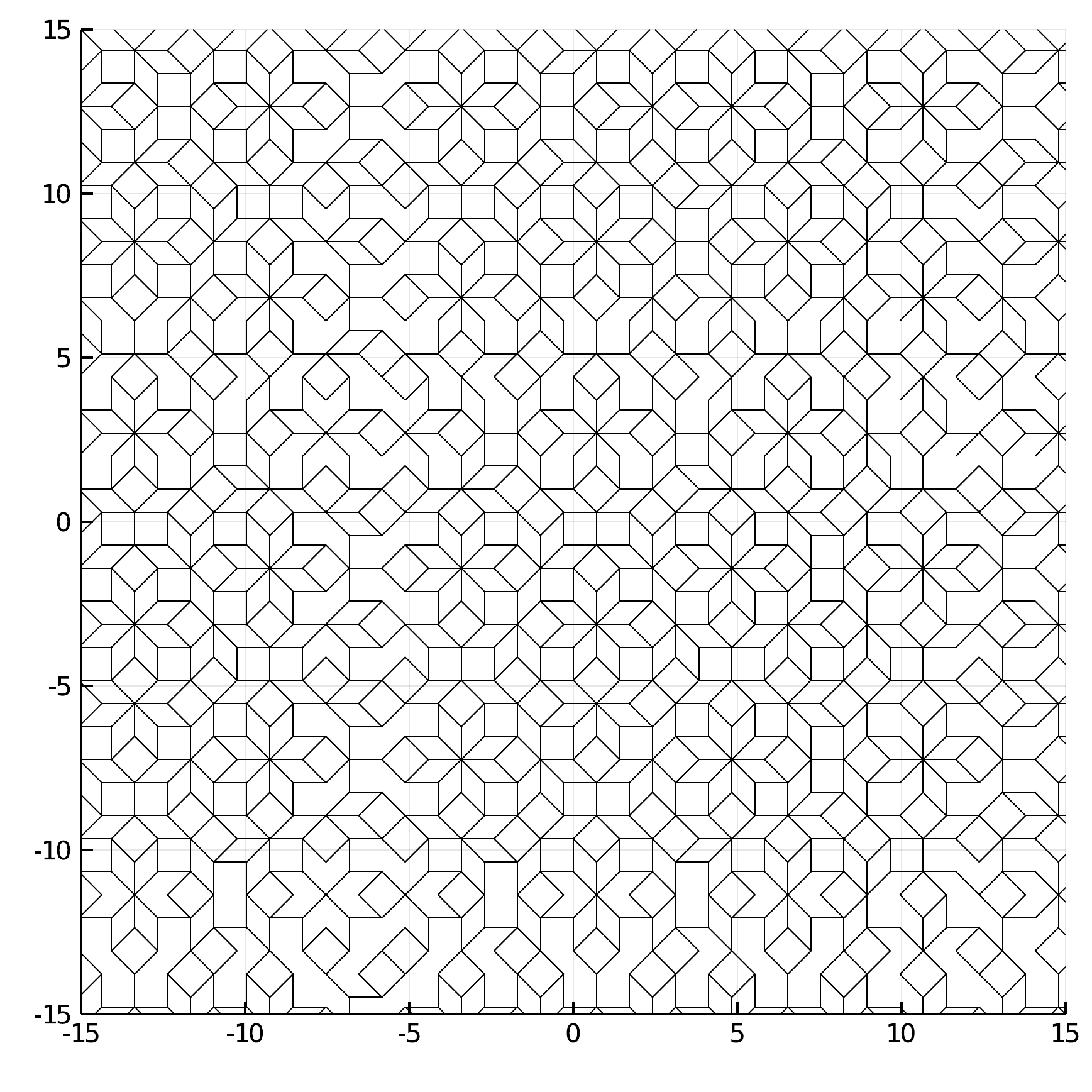}} 
        \end{tabular}
    \end{center}
    \caption{Ammann-Beenker lattices obtained by the dual grid method with different $\vec{\gamma}^{\rm T}$. (a): $\vec{\gamma}^{\rm T}= (0.1,0.14,-0.23,-\sum_{\mu=0}^2\gamma_{\mu})$ and (b): $\vec{\gamma}^{\rm T}= (-0.21,0.29,0.98, -\sum_{\mu=0}^2\gamma_{\mu})$. 
    The total numbers of lattice sites are 54579 and 54607, respectively. 
    \label{fig:ab}
     }
    \end{figure}

We adopt the so-called dual-grid method to generate the quasicrystals.
It is known that a $D$-dimensional quasicrystal can be obtained by the projection of a particularly cut slice of the $M$-dimensional euclidian hyper-lattice onto a $D$-dimensional plane.
As shown in Figs.~\ref{fig:pen}-\ref{fig:ab}, we can easily generate different patterns of Penrose or Ammann-Beenker lattices with different $\vec{\gamma}$, which determines a position of cut slice of the $M$-dimensional euclidian hyper-lattice. 
In other words, one can reproduce same quasicrystal structure with the use of same $\vec{\gamma}$.
The detail of the dual-grid method is shown in Appendix \ref{sec:dual}.

\subsection{Numerical approach for large-scale superconductors: LK-RSCG method with sparse modeling approach}
To consider superconducting quasicrystals, we have to solve the BdG equations (\ref{eq:bdg}) in real space with solving the gap equation Eq.~(\ref{eq:delta}) self-consistently. 
It is very hard to diagonalize the BdG Hamiltonian matrix in large quasicrystals, since the computational complexity to diagonalize the Hamiltonian matrix is ${\cal O}(N^3)$. 
Recently, we have proposed the localized Krylov-Bogoliubov-de Gennes method (LK-BdG), whose computational complexities are ${\cal O}(N)$ for self-consistent calculations and ${\cal O}(1)$ for calculating the local quantities such as the local density of states\cite{NagaiLK}. 
In the previous paper, we proposed the Chebyshev polynomial method and the Lanczos method as the applications of the LK-BdG method. 
In this paper, we adopt the reduced-shifted conjugate-gradient (RSCG) method to control the numerical accuracy\cite{NagaiRSCG}, which is also one of the applications of LK-BdG method.
The RSCG method uses the fact that the mean-fields are calculated by the solutions of linear equations as 
\begin{align}
\langle c_{i\downarrow} c_{i \uparrow}\rangle &= T \sum_{n=-\infty}^{\infty} \Vec{e}(i)^T \Vec{x}(i,\omega_n), \label{eq:x}
\end{align}
where the solutions $\Vec{x}(i,\omega_n)$ are obtained by solving the following linear equations: 
\begin{align}
(i \omega_n \hat{I} - \hat{H}) \Vec{x}(i,\omega_n) &= \Vec{h}(i). \label{eq:linear}
\end{align}
Here, $\omega_n \equiv \pi T (2n+1) $ denotes the Fermion Matsubara frequency and  $2N$-component unit-vectors $\Vec{e}(i)$ and $\Vec{h}(j)$ $(1 \leq i \leq N)$ are defined as
\begin{align}
[\Vec{e}(i)]_{\gamma} = \delta_{i,\gamma}, \: \: \: [\Vec{h}(i)]_{\gamma} = \delta_{i+N,\gamma},
\end{align}
The detail is shown in Appendix \ref{sec:rscg}. 

Let us introduce the Localized-Krylov subspace RSCG (LK-RSCG). 
Equation (\ref{eq:linear}) with different frequencies can be solved simultaneously with the use of the RSCG\cite{NagaiRSCG}.
Focusing on the fact that the vectors $\Vec{e}(i)$ and $\Vec{h}(i)$ are localized in real space, we show that the $k$-th order Krylov subspace generated by the Hamiltonian matrix given as\cite{NagaiLK} 
\begin{align}
{\cal K}_k(\hat{H},\Vec{b}) = {\rm span} \: (\Vec{b},\hat{H}\Vec{b},\hat{H}^2 \Vec{b},\cdots,\hat{H}^{k-1} \Vec{b}),
\end{align}
is localized in real space. 
The computational complexity of the matrix-vector products in the RSCG method ${\cal O}(N)$ is replaced with ${\cal O}(k^d)$. 
Here, $d$ is a spatial dimension\cite{FurukawaMotome,NagaiLK}.

In addition, to reduce the computational complexity, we introduce the recently-developed sparse-modeling (SpM) approach for a Green's function\cite{Chikano,NagaiJPSJ, Otsuki,Li,Wang}. 
The Green's function in Nambu-Gor'kov space can be defined as the solution of the Nambu-Gor'kov equation: 
\begin{align}
    (i \omega_n \hat{I} - \hat{H}) \hat{G}(i \omega_n) = \hat{1}.
\end{align}
With the use of the intermediate representation (IR) basis, the matrix element of the Green's function is written as 
\begin{align}
    G_{ij}(i \omega_n) &= \sum_{l=0}^{N_{\rm IR}-1} G_{l,ij} U_l(i \omega_n),
\end{align}
where $U_l(i \omega_n)$ is a basis function of the IR basis and  $N_{\rm IR}$ is a number of the basis functions(See, Appendix \ref{sec:spm}).
Here, $G_{ij}$ is the element of the matrix-valued Green's function $G_{ij} = [\hat{G}]_{ij}$. 
The superconducting mean-fields are given as 
\begin{align}
\langle c_{i\downarrow} c_{i \uparrow}\rangle &=\sum_{l=0}^{N_{\rm IR}-1} U_l(\beta) \sum_{k=0}^{N_{\rm IR}-1}  [\Vec{U}^{-1}]_{lk} \Vec{e}(i)^T \Vec{x}(i,\omega_k).
\end{align}
Although the size of $N_{\rm IR}$ depends on a cutoff parameter in the IR basis, $N_{\rm IR}$ is small around 10-100. 
With the use of the RSCG method and IR basis, we can calculate Eq.~(\ref{eq:x}) with high accuracy. 
With the use of the SpM, 
the total complexity for self-consistent calculation is ${\cal O}(m N_{\rm IR}) + {\cal O}(N m)$. 
The reduction of the computational complexity is summarized in Table \ref{table:1}. 

\begin{table}[t]
\caption{Reduction of computational complexity. Here, $N$, $N_{\rm IR}$, $n_{\rm cut}$, $m$ are the matrix size of the Hamiltonian, the number of the intermediate basis, the number of the Matsubara summations, and the number of the iteration steps for the RSCG. 
${\cal O}(N^3)$ in the table is the computational cost for the full diagonalization method of the Hamiltonian matrix.}
\label{table:1}
\begin{ruledtabular}
\begin{tabular}{ccccc}
Methods &
Computational complexity  \\
SpM approach & ${\cal O}(n_{\rm cut}) \rightarrow {\cal O}(N_{\rm IR})$  \\
LK matrix-vector operation & ${\cal O}(N) \rightarrow {\cal O}(1)$ \\
Total& ${\cal O}(N^3) \rightarrow {\cal O}(m N_{\rm IR}) + {\cal O}(N m)$  \\
\end{tabular}
\end{ruledtabular} 
\end{table}

\section{Results}\label{sec:results}
We solve the gap equations self-consistently in the Penrose and Ammann-Beenker quasicrystals. 
We consider the temperature $T = 10^{-3} t$ and the cutoff frequency $\omega_{\rm max} = 10t$, where the number of the IR basis $N_{\rm IR}$ is 104.

\subsection{Basic properties of superconducting quasicrystal without vortices}
Superconductivity without vortices on the Penrose and Ammann-Beenker tiling has been studied by several groups, respectively\cite{Araujo,Takemori2020}.  
According to the previous studies, there are two important properties of superconductivity on quasicrystals.

\begin{figure}[t]
    %%%%--- I comment out figure regions
    %\vspace{50mm}
    \begin{center}
    \resizebox{0.7 \columnwidth}{!}{\includegraphics{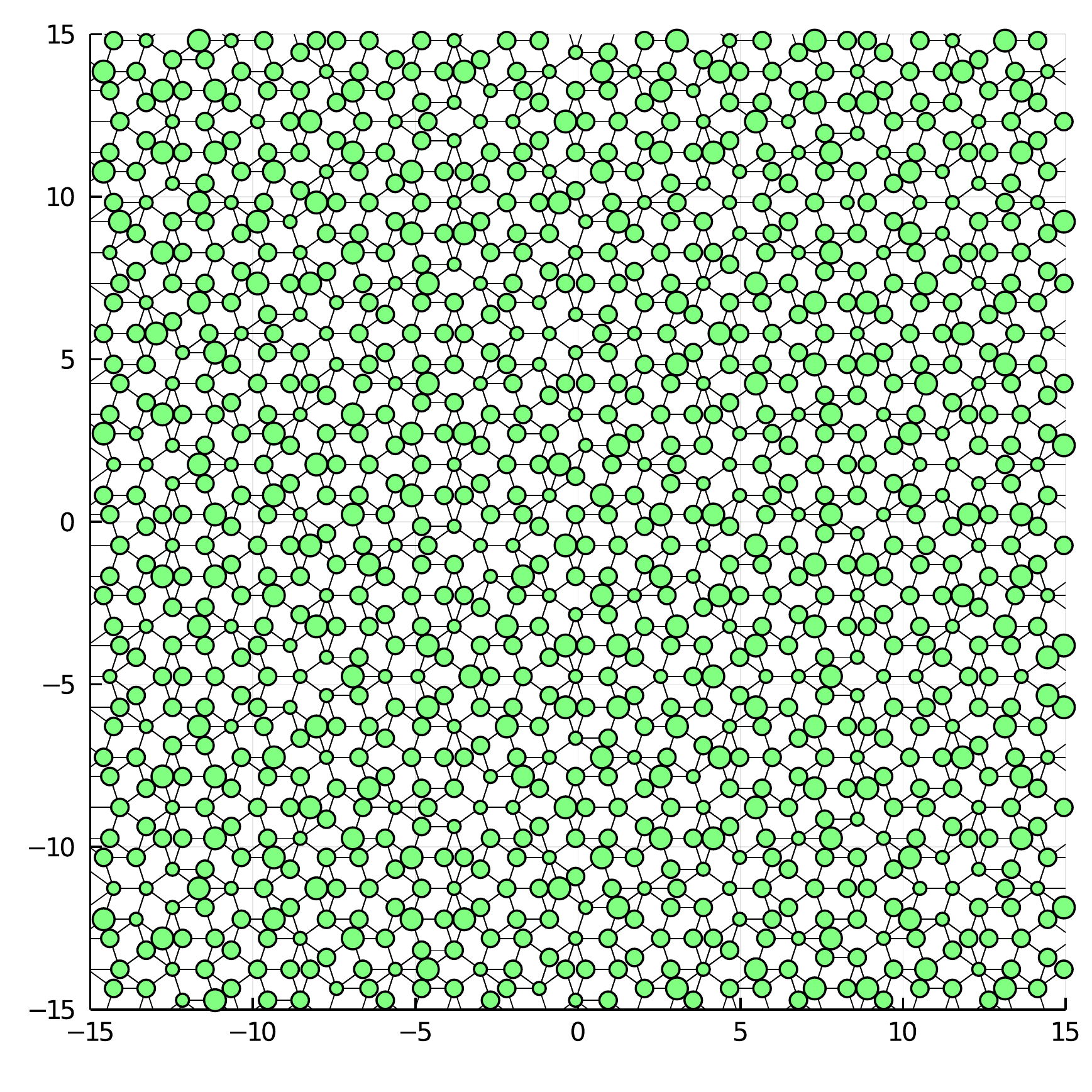}} 
    \end{center}
     \begin{center}
    \resizebox{0.9 \columnwidth}{!}{\includegraphics{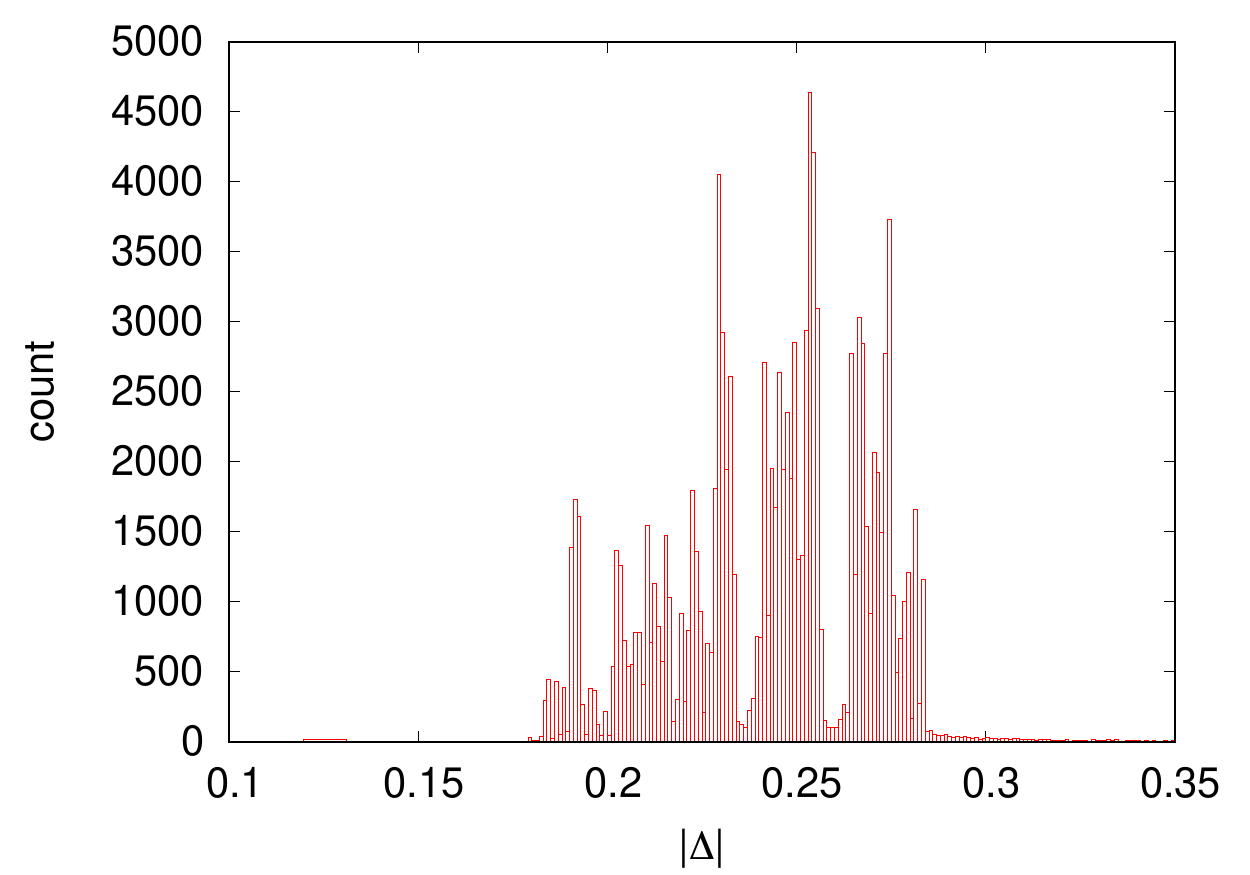}} 
    \end{center}
    \caption{
    (Color online) (Upper panel) Superconducting order parameter around the origin in a system characterized by $\vec{\gamma}^{\rm T}_A = (0.1,0.7,-0.98,0.43,-\sum_{\mu=0}^3\gamma_{\mu})$ . 
    The size of the circles is proportional to the amplitude of the order parameter. 
    (Lower panel) Superconducting order parameter distribution. The unit of the order parameter is the hopping amplitude $t$. 
        \label{fig:withoutvortex}
     }
    \end{figure}

The first is a non-uniform distribution of the superconducting order parameter. 
Even if the onsite pairing interaction does not depend on a position in real space, the superconducting order parameter defined by Eq.~(\ref{eq:delta}) is inhomogeneous in real space. 
As shown in Fig.~\ref{fig:withoutvortex}, we confirm that similar inhomogeneity occurs in our parameter set of the Penrose tiling. 
Here,  we consider the chemical potential $\mu = -t$ and the onsite pairing interaction $U=-2t$.
We choose $\vec{\gamma}_A= (0.1,0.7,-0.98,0.43,-\sum_{\mu=0}^3\gamma_{\mu})$ and the radius of the system is $172a$, where $a$ is the bond length of the lattice and the total number of the lattice sites is 119026. 
%The length unit is the bond length of the Penrose lattice. 
%As shown in Fig.~\ref{fig:withoutvortex}, the superconducting order parameter is inhomogeneous. 
In Fig.~\ref{fig:withoutvortex}(a), the spatial average of the amplitude of order parameter is $\sim 0.24344t$. 
We also confirm that the spatial average of the amplitude of order parameter does not depend on the parameter $\gamma$.

The second is that many thermodynamic properties in superconducting quasicrystal are almost indistinguishable from ones in conventional BCS superconductors. 
Ara\'{u}jo and Andrade found no evidence of superconductivity islands although the superconducting order parameter is inhomogeneous in Ammann-Beenker quasicrystal and they concluded that quasicrystals are prone to display conventional BCS-like superconductivity\cite{Araujo}. 
Takemori {\it et al.} calculated the local density of states, specific heat, I-V characteristics in Penrose quasicrystal\cite{Takemori2020}.
Although they claimed that there are difference between Penrose superconducting quasicrystal and conventional BCS superconductors, the difference is not very large, which might not be detected in experiments.  
The spectrum gap in the local density of states is almost uniform in real space where the order parameter is not uniform, which suggests that the STM/STS measurements can not detect the inhomogeneity of the superconducting order parameter. 
There is the jump of the specific heat at the critical temperature and the amplitude of the jump $\Delta C$ is relatively smaller than the value in conventional BCS theory. 
However, in real materials, the jump $\Delta C$ depends on materials even if the superconductivity is explained by the BCS theory. 
In addition, according to the previous result on Ammann-Beenker quasicrystal\cite{Araujo}, the increase of the critical temperature due to a multifractal nature of the electronic state is not found in this system, as is expected on disordered systems close to the Anderson metal-insulator transition.

It is difficult to detect the inhomonogeneous phenomena in experiments of superconducting quasicrystals without magnetic fields.
%The atomic scale inhomogeneity of the superconducting order parameter can not be directly measured since the local density of states is almost uniform. 
Therefore, we need a vortex as a source of the quasiparticle excitation. 

\subsection{Electronic structure with a vortex}
Let us consider systems with a vortex. 
We study superconductivity in Penrose and Ammann-Beenker quasicrystals in the type II limit (the magnetic penetration depth $\lambda \rightarrow \infty$). 
As the initial state, we locate a vortex at a center in each system. 
After solving gap equations self-consistently, a vortex moves to find the position where the free energy becomes minimum. 
We show results about both Penrose and Amman-Beenker lattices in this section. 

\subsubsection{Penrose quasicrystal}
We consider four patterns of Penrose lattice, which are determined by the following $\vec{\gamma}$, respectively: 
\begin{align}
\vec{\gamma}^{\rm T}_A &=   (0.1,0.7,-0.98,0.43,-\sum_{\mu=0}^3\gamma_{\mu}), \\
\vec{\gamma}^{\rm T}_B &= (1/7,1/9,-3/4,-\sqrt{5}, -\sum_{\mu=0}^3\gamma_{\mu}), \\
\vec{\gamma}^{\rm T}_C &=  (0.1,0.2,0.3,0.4, -\sum_{\mu=0}^3\gamma_{\mu})), \\
\vec{\gamma}^{\rm T}_D &= (-0.4,-0.13,2.4,0.89, -\sum_{\mu=0}^3\gamma_{\mu}).
\end{align}
We consider the chemical potential $\mu = -t$ and the onsite pairing interaction $U=-2t$.

\begin{figure}[t]
    %%%%--- I comment out figure regions
    %\vspace{50mm}
    \begin{center}
         \begin{tabular}{p{ 0.5 \columnwidth} p{0.5 \columnwidth}}%  p{28mm}}
          (a) \resizebox{0.5 \columnwidth}{!}{\includegraphics{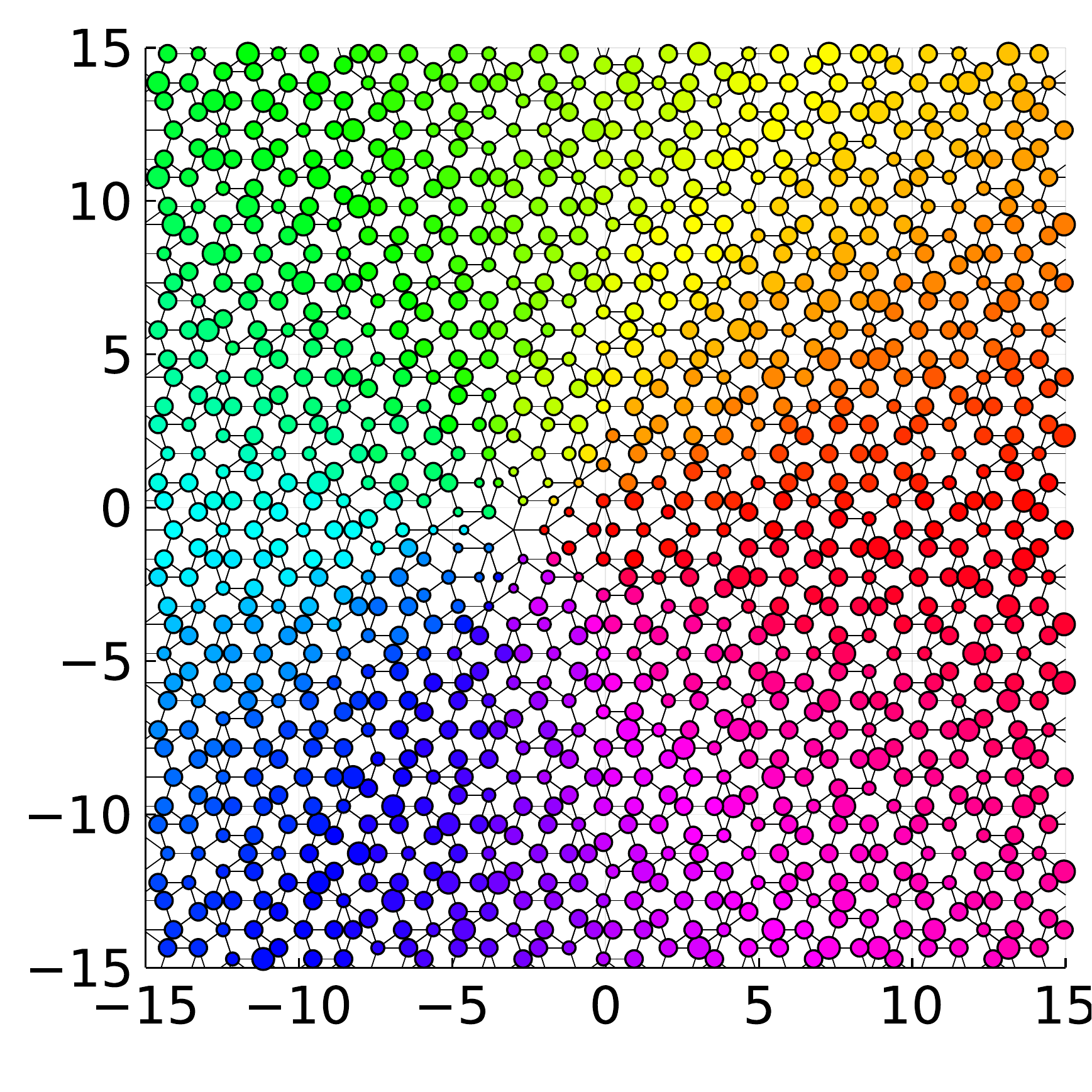}} 
          &
          (b) \resizebox{0.5 \columnwidth}{!}{\includegraphics{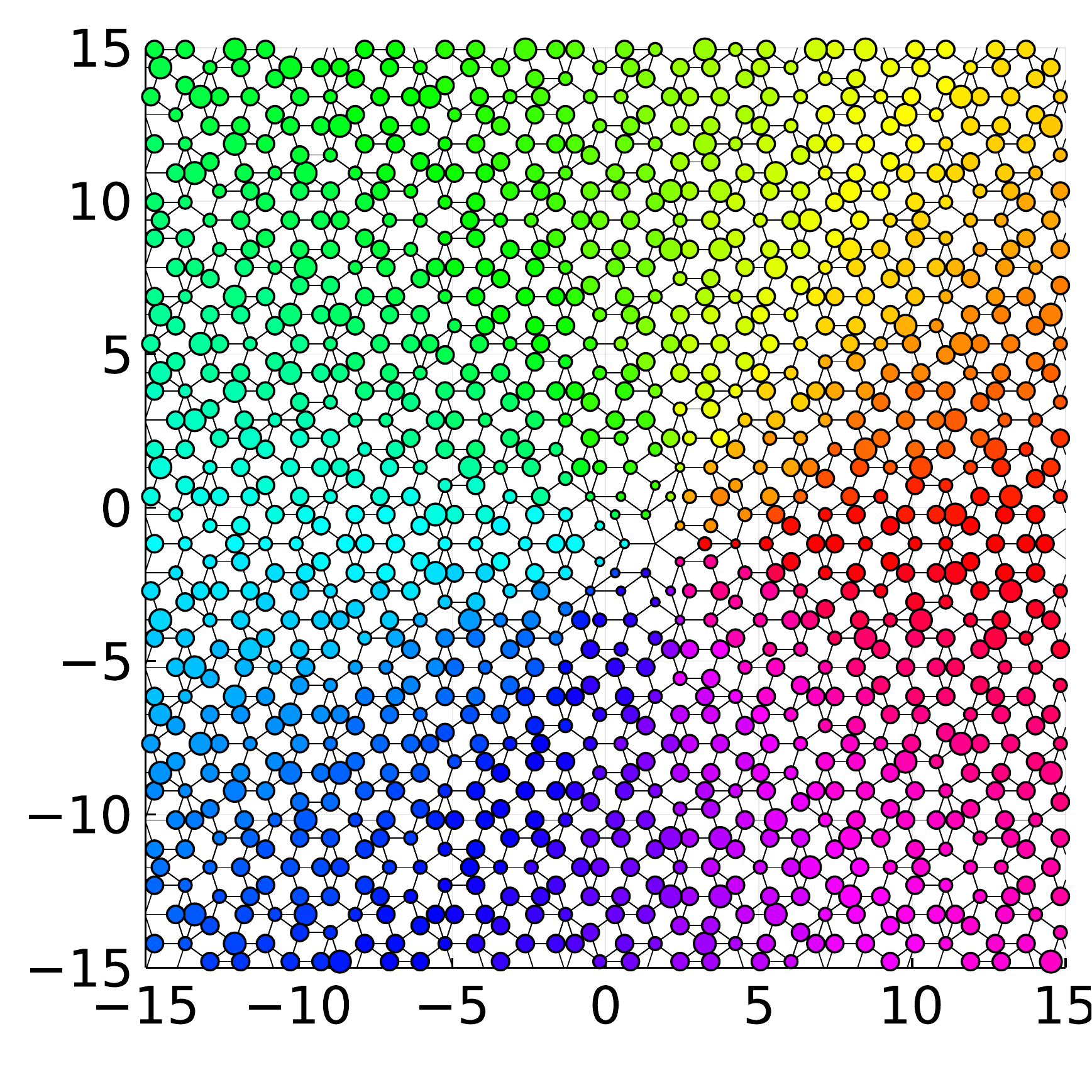}} \\
          (c) \resizebox{0.5 \columnwidth}{!}{\includegraphics{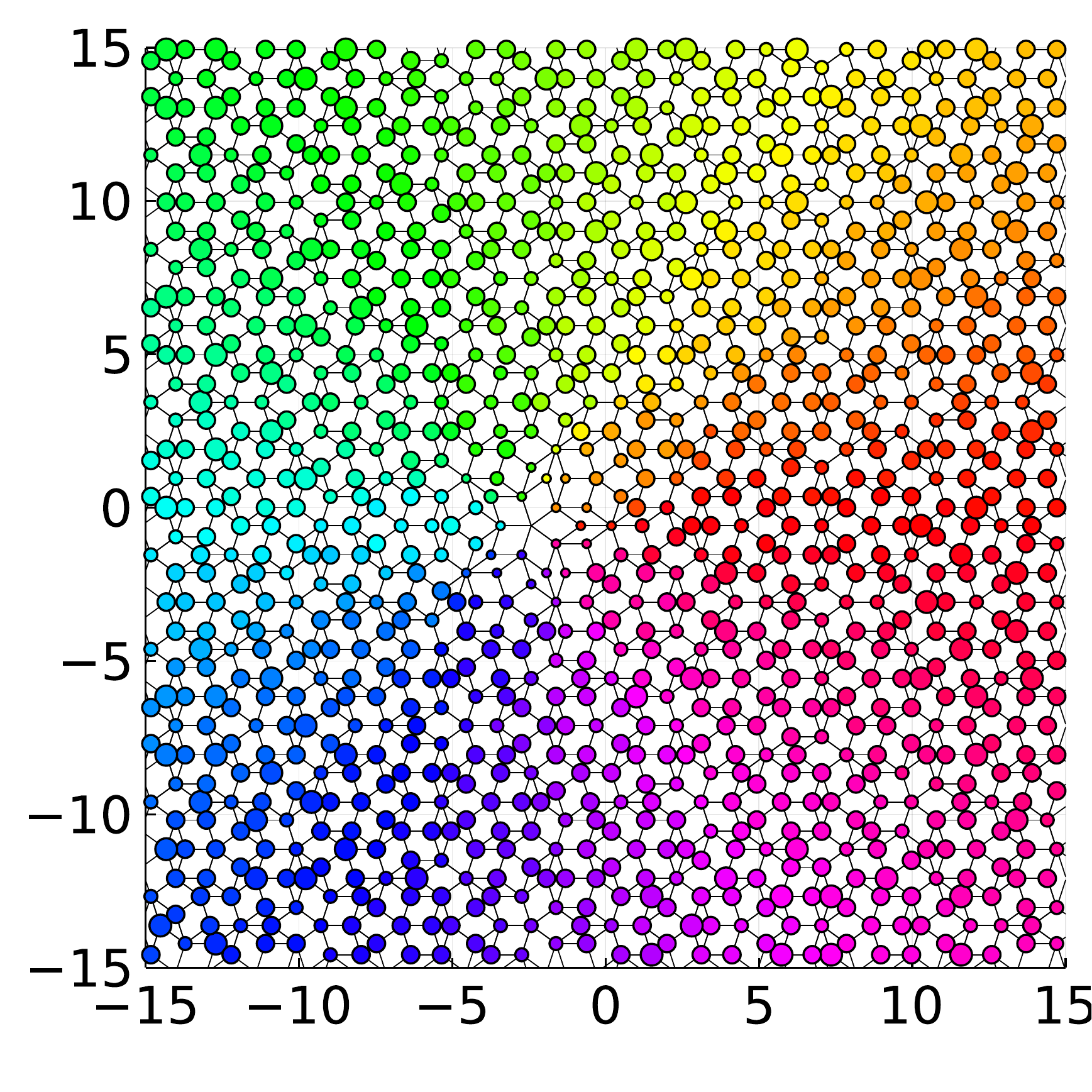}} 
          &
          (d) \resizebox{0.5 \columnwidth}{!}{\includegraphics{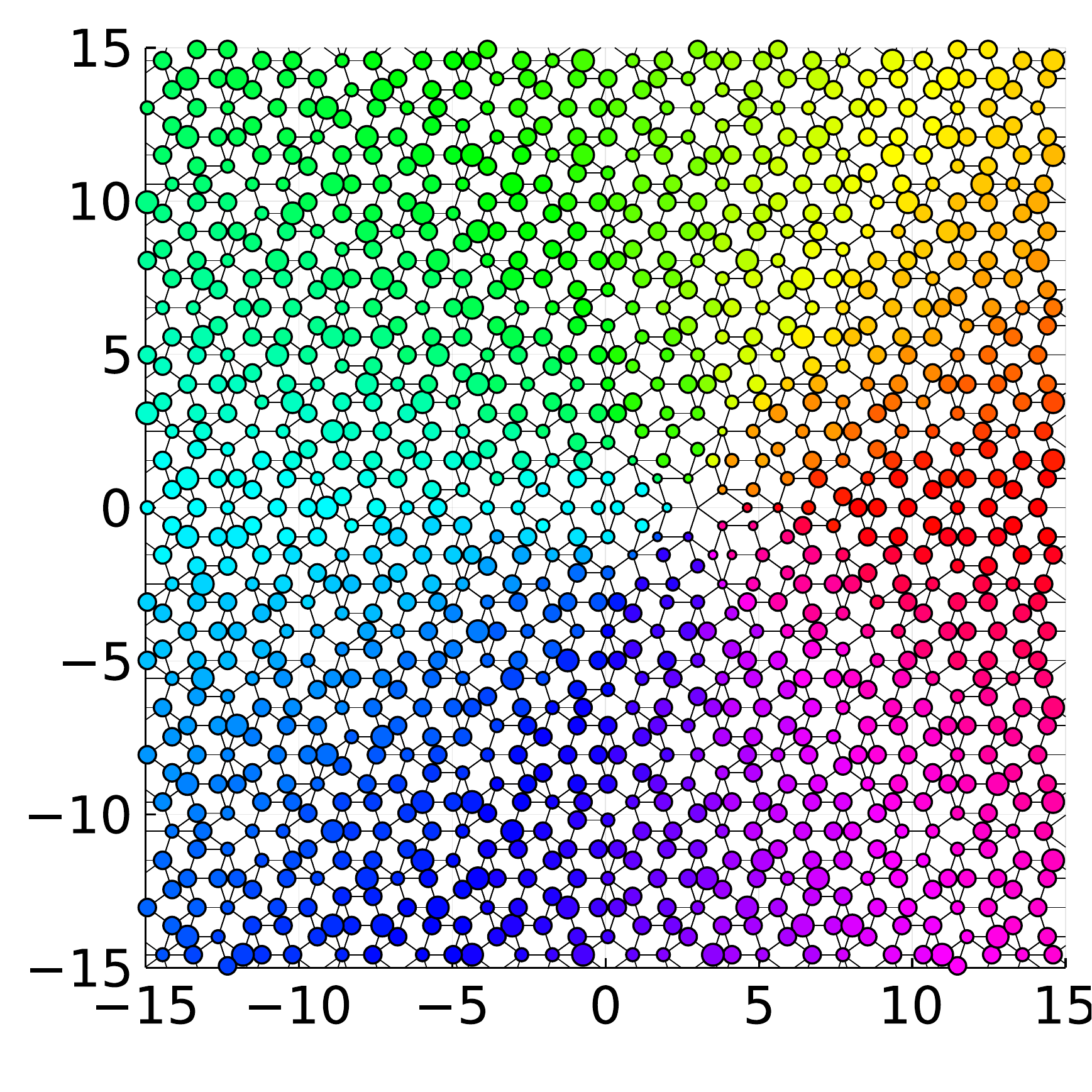}} 
        \end{tabular}
         \resizebox{0.3 \columnwidth}{!}{\includegraphics{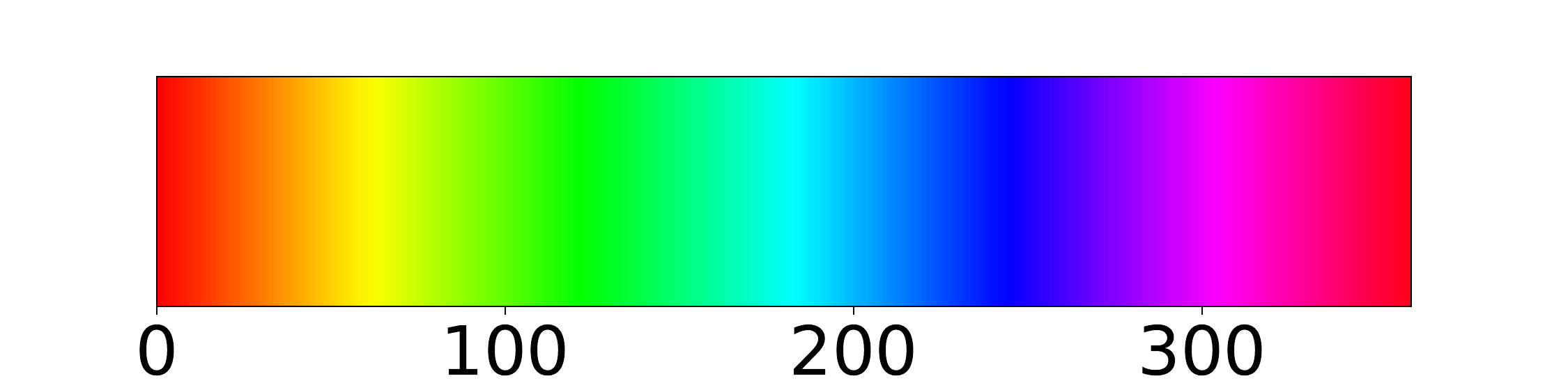}}
    \end{center}
    \caption{(Color)
    Superconducting order parameter with different $\vec{\gamma}^{\rm T}$ in superconducting Penrose quasicrystals. 
    The size of the circles is proportional to the amplitude of the order parameter. 
    The color represents the phase of the order parameter. 
    (a): $\vec{\gamma}^{\rm T}_A= (0.1,0.7,-0.98,0.43,-\sum_{\mu=0}^3\gamma_{\mu})$, (b): $\vec{\gamma}^{\rm T}_B= (1/7,1/9,-3/4,-\sqrt{5}, -\sum_{\mu=0}^3\gamma_{\mu})$, (c)$\vec{\gamma}^{\rm T}_C =  (0.1,0.2,0.3,0.4, -\sum_{\mu=0}^3\gamma_{\mu}))$ and (d)$\vec{\gamma}^{\rm T}_D = (-0.4,-0.13,2.4,0.89, -\sum_{\mu=0}^3\gamma_{\mu})$.
    \label{fig:penamp}
     }
    \end{figure}

After solving gap equations self-consistently,  a vortex is not located at the center as shown in Fig.~\ref{fig:penamp}. 
The position of the vortex is different in each system but the local lattice structure around a vortex looks similar to each other. 
Since the system with the converged gap distribution has minimum free energy, the free energy depends on the vortex position on the Penrose quasicrystal. 
Therefore, we claim that the intrinsic vortex pinning occurs on superconducting quasicrystals. 

    \begin{figure}[th]
    %%%%--- I comment out figure regions
    %\vspace{50mm}
    \begin{center}
    \resizebox{0.9 \columnwidth}{!}{\includegraphics{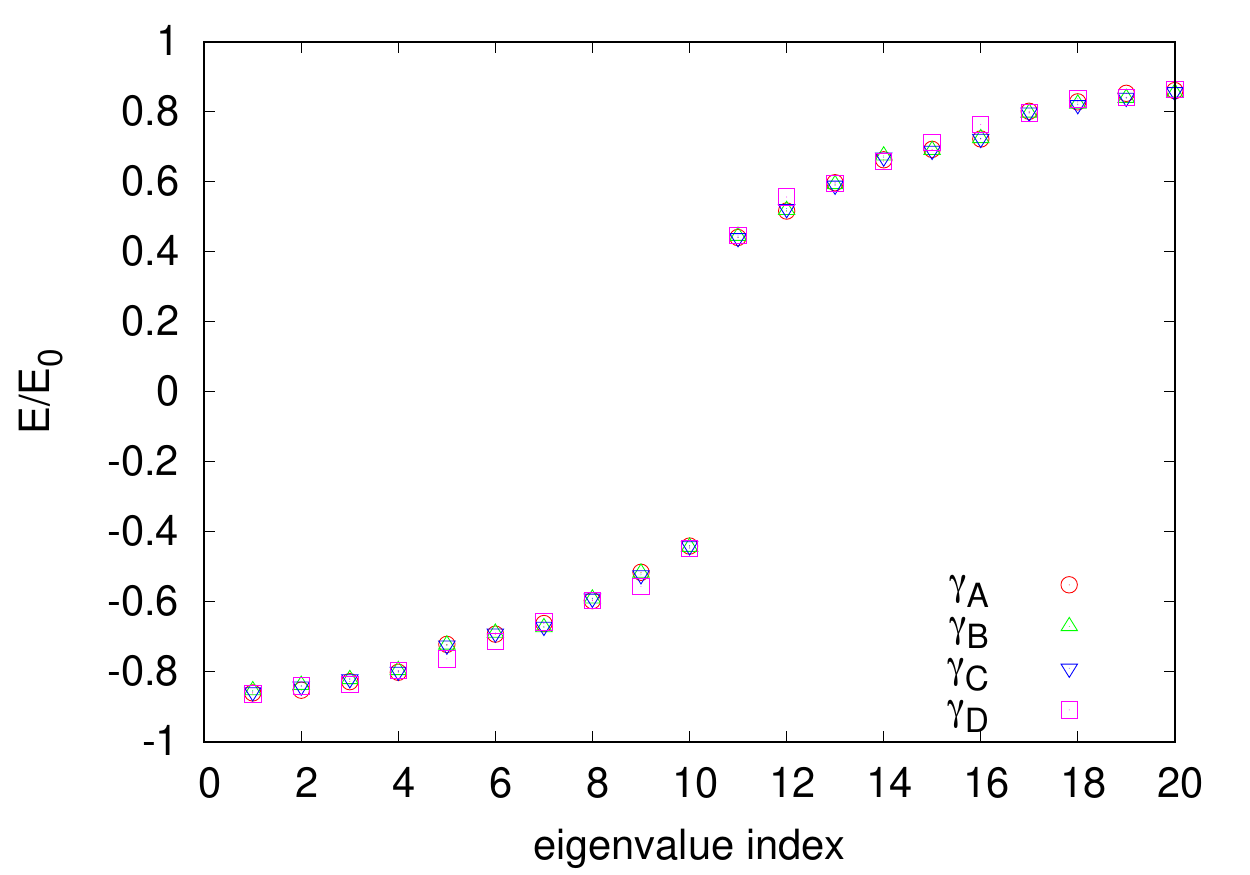}} 
    \end{center}
    \caption{(Color online)
    Eigenvalues in different systems with a vortex in the Penrose quasicrystal. 
    $E_0 = 0.243t$ %44170278384056t
    is the absolute minimum of the eigenvalues in the system without vortex, where the total number of the lattice sites is 119026. 
    \label{fig:eigenvalue}
     }
    \end{figure}

Vortex bound states in superconductors have rich information about a superconducting order parameter. 
We calculate the low-energy eigenvalues of the BdG Hamiltonian in four systems, originating from the bound states around a vortex core. 
As shown in Fig.~\ref{fig:eigenvalue}, we find that the low-energy eigenvalue distribution does not depend on systems. 
The lowest eigenvalue is around $0.5E_0$. 
Here, $E_0$ ($= 0.243t$) is the absolute minimum of the eigenvalues in the system without vortex,  where the total number of the lattice sites is 119026.
In conventional $s$-wave BCS superconductors, the energy levels are equally spaced, which is characterized by $\Delta^2/E_{\rm F}$, a ratio of the order parameter $\Delta$ and Fermi energy $E_{\rm F}$\cite{HayashiPRL}. 
However, we found that the energy of the first excited vortex bound state is much larger than the energy difference between the first and second bound states. 
This "quantum-limit" behavior might be observed by the STM/STS experiments. 
The reason of this behavior is explained in the discussion section. 

The size of the vortex core in superconducting Penrose lattice looks small in Fig.~\ref{fig:penamp}. 
This small core might be understand by a core shrinkage effect so-called the Kramer-Pesch effect\cite{Kramer-Pesch,HayashiPRL}, which occurs conventional superconductors. 
In conventional superconductors, the vortex core is characterized by the coherence length $v_{\rm F}/\Delta$, where $v_{\rm F}$ is the Fermi velocity. 
The Kramer-Pesch effect becomes large when the energy-level spacing of the bound states is large\cite{HayashiPRL}. 
We should note that the coherence length can not be determined with the use of the Fermi velocity in quasiperiodic systems, since the Fermi velocity is not defined in these systems. 
Therefore, the definition of the coherence length is not clear in superconducting quasicrystals. 

We should point out that the energy of the lowest vortex bound states is much larger than the bound energy in the system created by the inflation-deflation method in our previous paper\cite{NagaiLK}. 
We have reported that the energy of the vortex bound states is almost zero in the system\cite{NagaiLK}.
This difference comes from the position of the vortex. 
In the previous paper, the vortex can not move from a center because the vortex is located at a center of the system with a 10-fold symmetry. 
The minimum energy is determined by the local structure around a center. 
In the 21106-site 143806-site Penrose lattice systems  whose local lattice structure around a center is same, the minimum energy is close to zero. 
However, we have also reported that, in the 375971-site Penrose lattice whose local structure differs from that in 21106-site, the minimum energy level in a vortex core is larger. 
On the other hand, in systems created by dual-grid method in this paper, the vortex can move since there is no local 10-fold symmetry.
These result suggests that the free energy depends on the local lattice structure around a vortex, which is discussed later.

\subsubsection{Ammann-Beenker quasicrystal}
\begin{figure}[t]
    %%%%--- I comment out figure regions
    %\vspace{50mm} 
    \begin{center}
         \begin{tabular}{p{ 0.5 \columnwidth} p{0.5 \columnwidth}}%  p{28mm}}
          (a) \resizebox{0.5 \columnwidth}{!}{\includegraphics{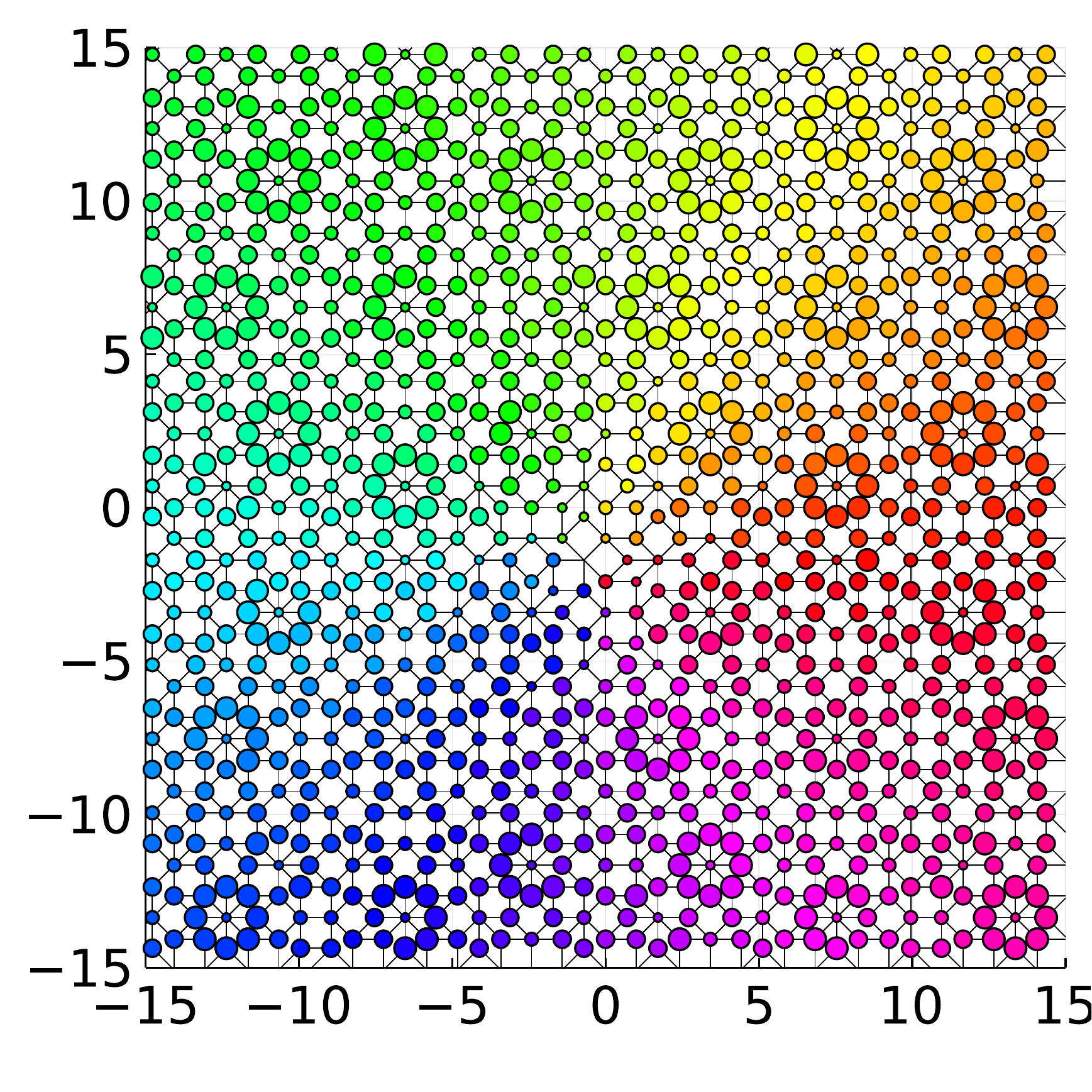}} 
          &
          (b) \resizebox{0.5 \columnwidth}{!}{\includegraphics{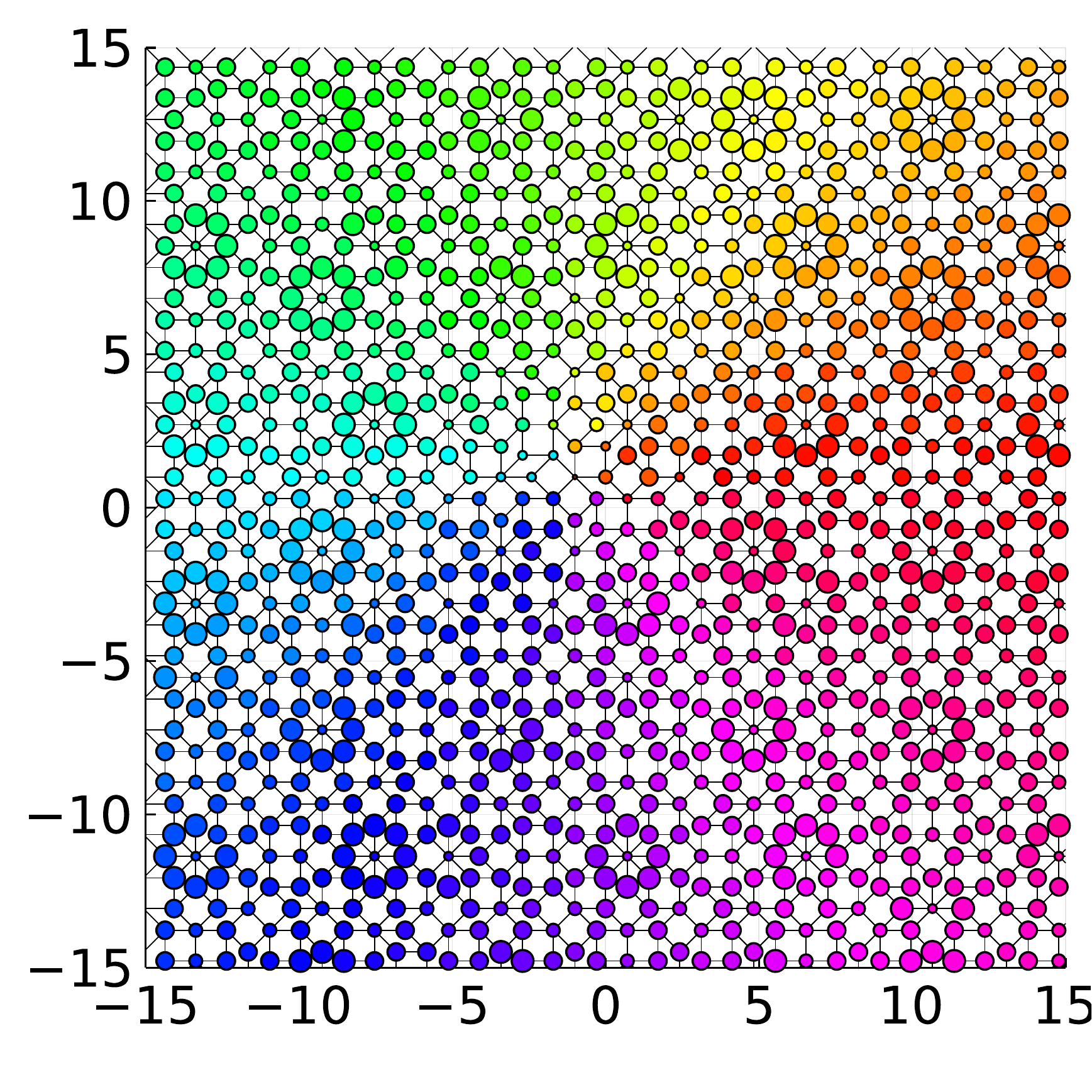}} 
        \end{tabular}
         \resizebox{0.3 \columnwidth}{!}{\includegraphics{colorbar.pdf}}
    \end{center}
    \caption{(Color) 
    Superconducting order parameter with different $\vec{\gamma}^{\rm T}$ in superconducting Ammann-Beenker quasicrystals. 
    The size of the circles is proportional to the amplitude of the order parameter. 
    The color represents the phase of the order parameter. (a): $\vec{\gamma}^{\rm T}= (0.1,0.14,-0.23,-\sum_{\mu=0}^2\gamma_{\mu})$ and (b): $\vec{\gamma}^{\rm T}= (-0.21,0.29,0.98, -\sum_{\mu=0}^2\gamma_{\mu})$.
    The mean value of the order parameter in the system without a vortex is $0.26593t$.
    \label{fig:abd}
     }
    \end{figure}

\begin{figure}[th]
    %%%%--- I comment out figure regions
    %\vspace{50mm}
    \begin{center}
    \resizebox{0.9 \columnwidth}{!}{\includegraphics{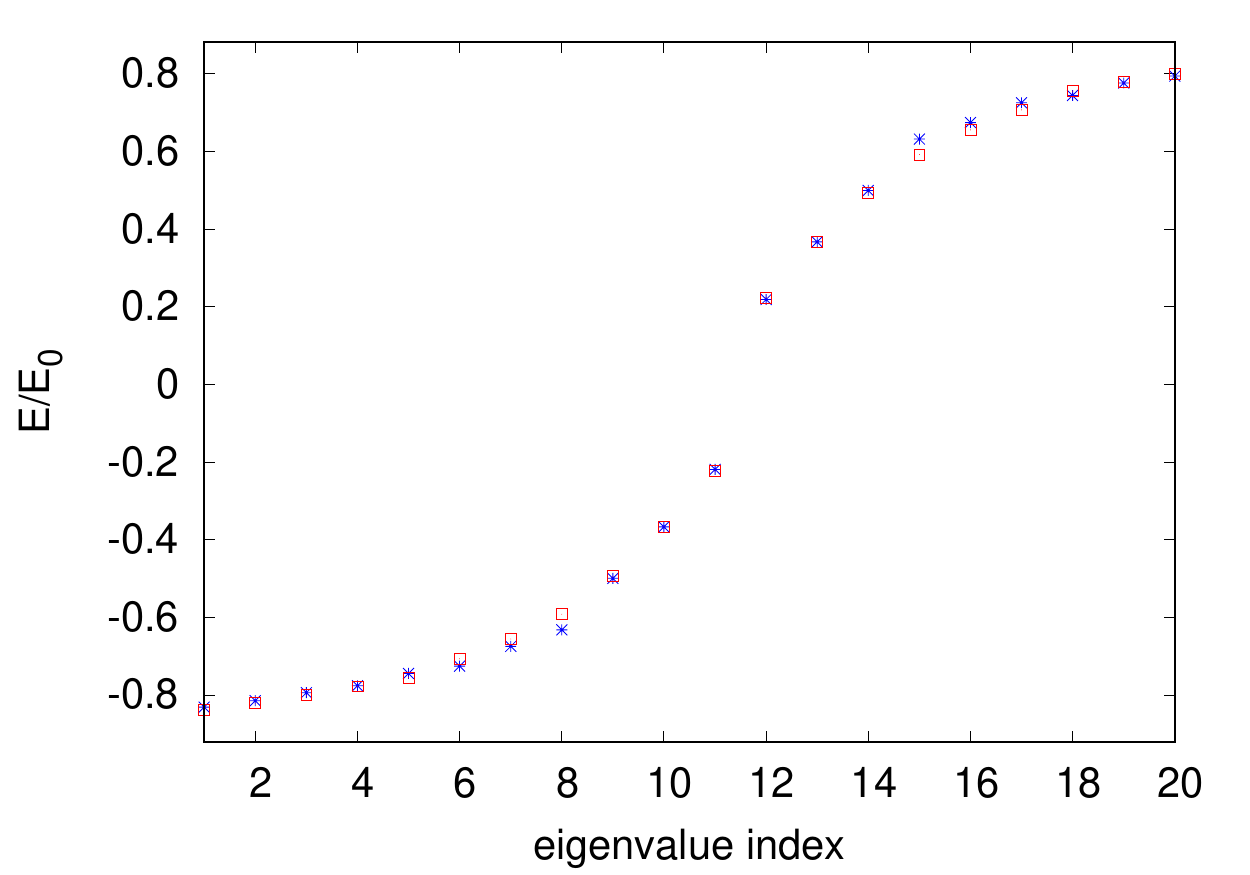}} 
    \end{center}
    \caption{(Color online)
    Eigenvalues in different systems with a vortex in Ammann-Beenker lattices. 
    $E_0 =0.2597t$
    %8888504892795t$
    is the absolute minimum of the eigenvalues in the system without vortex, where the total number of the lattice sites is 54604. 
    \label{fig:eigenvalue_ab}
     }
    \end{figure}

We consider the Ammann-Beenker quasicrystal to confirm whether an intrinsic vortex pinning occurs in another quasicrystal structures.
We consider two patterns of AB lattice, which are determined by the following $\vec{\gamma}$, respectively: 
\begin{align}
\vec{\gamma}^{\rm T}_A &=   (0.1,0.14,-0.23,-\sum_{\mu=0}^2\gamma_{\mu}), \\
\vec{\gamma}^{\rm T}_B &= (-0.21,0.29,0.98, -\sum_{\mu=0}^2\gamma_{\mu}).
\end{align}
In the system without a vortex in our parameter set, the mean value of the superconducting order parameter is $0.26593 t$. 
The basic property of the superconducting  Ammann-Beenker quasicrystal is similar to the Penrose one. 
We note that there is a paper discussing the system without a vortex\cite{Araujo}. 
For systems with a vortex, we initially put a vortex at a center and solve the gap equations. 
After solving gap equations self-consistently, a vortex is not located at the center as shown in Fig.~\ref{fig:abd}. 
We find that the local lattice structure near a vortex in two systems is similar to each other. 
We also calculate the low-energy eigenvalues of two systems as shown in Fig.~\ref{fig:eigenvalue_ab}. 
One can clearly find that the bound state energy of a vortex is same in two systems. 
Therefore, the vortex in AB lattice is also pinned.

%The initial location of a vortex is a center of the system. 
%Figure \ref{fig:penamp} shows that a vortex is not located at the center. 

%We also confirm that the absolute minimum eigenvalue in a system () without vortex is $\sim 0.244$

\section{Discussions}\label{sec:discussions}
\subsection{Mechanism of the intrinsic vortex pinning in superconducting quasicrystals}
We found that the intrinsic vortex pinning occurs in superconducting quasicrystals. 
In the mean-field level, solving the gap equation (\ref{eq:delta}) means finding the distribution of the superconducting order parameter where the thermodynamic potential becomes minimum. 
The vortex in superconducting Penrose and Ammann-Beenker  tight-binding models is pinned, where the energies of vortex bound states are large. 
We explain the reason as follows. 
We calculate the low-energy eigenvalues of the BdG Hamiltonian in four systems, which reflects the bound states around a vortex core. 
As shown in Fig.~\ref{fig:eigenvalue}, we find that the low-energy eigenvalue distribution does not depend on systems, which suggests that a vortex is pinned where a thermodynamic potential becomes minimum. 
The thermodynamic potential $ \Omega$ in the BdG framework is defined as \cite{Kosztin,Hosseini,NagaiPRB2017} 
\begin{align}
    \Omega &= -T \sum_{\gamma = 1}^{2N} \ln \left[ 1 + \exp \left(\frac{E_{\gamma}}{T} \right) \right]  - \sum_{i} \frac{|\Delta_i |^2}{U}.
\end{align}
Since the temperature is very low ($T = 10^{-3} t$), $ \Omega$ is expressed as\cite{NagaiPRB2017} 
\begin{align}
\Omega \sim - \sum_{\gamma=1}^{2N} E_{\gamma} \theta(E_{\gamma}) - \sum_{i} \frac{|\Delta_i |^2}{U}, \label{eq:energy}
\end{align}
which is equivalent to the internal energy. 
Here $\theta(x)$ is the Heaviside step function. 
The first term in Eq.~(\ref{eq:energy}) decreases when the bound state energy becomes larger. 
The second term is usually small since the system satisfies the relation $|\Delta_i |^2/|U| < E_0$. 
Here, $E_0$ is the absolute minimum of the eigenvalues in the system without vortex. 
Equation (\ref{eq:energy}) shows that lower energy bound state less contributes the internal energy. 
In our previous paper\cite{NagaiLK}, we showed the minimum of the bound states depends on the local lattice structure around a vortex center in the Penrose quasicrystals.
%In superconducting quasicrystals, the order parameter is not uniform. 
Therefore, a vortex moves during self-consistent iteration loops and is pinned at the position where the total energy becomes minimum on superconducting quasicrystal.  

We claim that this intrinsic vortex pinning occurs in three-dimensional realistic materials, since the vortex pinning originates from the inhomogeneous superconducting order parameter. 
In general, the superconducting order parameter is defined as $\Delta({\bm R},{\bm r})$, where ${\bm R}$ and ${\bm r}$ are  the center-of-mass and the relative coordinates, respectively. 
In superconductors with the translational symmetry, the superconducting order parameter becomes $\Delta({\bm R},{\bm r}) = \Delta({\bm r})$. 
In quasicrystals, there is no translational symmetry.
The Cooper pairs in quasicrystals can not be characterized by single momenta.  
Then, fourier-transformed order parameter $\Delta({\bm k}_1,{\bm k}_2)$ becomes complicated function in momentum space\cite{SakaiPRB}.
In real space, there should be the center-of-mass dependence $\Delta({\bm R},{\bm r})$.
Therefore, the inhomogeneous superconducting order parameter is the intrinsic effect in superconducting quasicrystal in any dimension.

\subsection{Position of vortices on superconducting quasicrystals}
Penrose lattice has eight types of vertices with nomenclature of de Bruijn \cite{Bruijn1981part1,Bruijn1981part2,Koga2017}. 
In our parameter set of the Penrose superconductors, as shown in Fig.~\ref{fig:penamp}, the vortex is pinned at "S5" vertex, which has a local five-fold rotational symmetry. 
We also confirm that vortices are pinned at these vertices even if there are several vortices in systems. 
In the system without a vortex, the superconducting order parameter becomes minimum at this S5 vertex, and it becomes large at the other five-fold symmetric vertex so-called "S" vertex, 
as shown in Fig.~\ref{fig:withoutvortex}. 
We have studied the electronic structure in the system where the vortex is located at "S" vertex in our previous paper\cite{NagaiLK}.
As shown in Fig.2(d) in Ref.~\onlinecite{NagaiLK}, the vortex has near-zero energy bound states. 
Since near-zero bound states less contribute the thermodynamic potential given as Eq.~(\ref{eq:energy}), the vortex is not pinned at "S" vertex. 
According to the paper by Sakai {\it et al.}\cite{Sakai2022}, S and S5 vertices constitute the superlattice as shown in Fig.~9 in Ref.~\onlinecite{Sakai2022}. 
Since the vortex is pinned at S5 vertex not at S vertex, vortex lattice does not constitute the  Penrose-like superlattice in systems with our parameter set.

Vortex lattice structure becomes unique in superconducting quasicrystal, although we do not confirm that a vortex is always pinned at S5 vertex in all parameter region of the Penrose superconducting tight-binding model. 
The important point is that the amplitude of the superconducting order parameter depends on the local lattice structure of the quasicrystals without magnetic fields. 
Our calculation suggests that the site that has relatively smaller gap amplitude can be pinning site. 
Since the vortex suppresses the superconducting order parameter around a center of the vortex, it might be better to make the amount of the suppression small in terms of the total minimization of the thermodynamic potential. 

We should note that the vortex lattice structure is determined not only by the pinning site distributions, but also by the interaction between vortices\cite{Reichhardt_2016}. 
In a conventional superconductor, vortices form the Abrikosov triangular lattice. 
What kind of vortex structure occurs on quasicrystaline pinning potentials depends on a cooperation or competition between the pinning potential and vortex-vortex interactions\cite{PhysRevLett.95.177007,PhysRevLett.106.060603}. 
For example, the critical current on systems quasiperiodic pinning arrays is discussed\cite{PhysRevLett.95.177007}. 
A pattern formed by interacting particles on a quasiperiodic potentials has been studied also in colloids\cite{Mikhael,PhysRevLett.101.218302,PhysRevLett.109.058301}. 
For example, Archimedian like ordering is discussed, which occurs as a compromise between the quasicrystalline ordering and triangular ordering\cite{Mikhael,PhysRevLett.109.058301}.
A phason dynamics of vortices might be interesting in vortex systems, which has been discussed in colloids\cite{PhysRevLett.108.218301,Kromer}.
Here, the phason is a characteristic behavior in quasiperiodic systems.

\subsection{Experiments to detect vortex pinning}
If STM/STS or SQUID measurements can observe a vortex pinning in clean quasicrystals, this becomes evidence of the inhomogeneous superconductivity in quasicrystals. 
Other experiments influenced by a pinning effect also become tools to detect this inhomogeneity of superconductivity. 
For example, the critical current should depend on the pinning effect\cite{PhysRevLett.95.177007}. 

We consider clean and perfect quasicrystals in this paper.  
There can be kinds of extrinsic pinning such as voids and twins in realistic materials. 
In addition, in quasicrystals, there can be so-called phason strain, which is a characteristic dislocation in quasicrystal\cite{Socolar,YamamotoXray}.  
This dislocation can be also become a source of pinning potential. 
The actual pinning site is determined by the competition between intrinsic and extrinsic effect. 
We suggest that the intrinsic pinning would be the most prominent and would dominate over other extrinsic pinning effects near critical temperature.

%The initial location of a vortex is a center of the system. 
%Figure \ref{fig:penamp} shows that a vortex is not located at the center. 

%We also confirm that the absolute minimum eigenvalue in a system () without vortex is $\sim 0.244$

\section{Summary}\label{sec:summary}
We showed that an intrinsic vortex pinning due to the inhomogeneous superconducting order parameter occurs in superconducting quasicrystals.
We confirmed that the intrinsic vortex pinning occurs in systems with several vortices in both Penrose and AB quasicrystals.
We proposed the new method to solve the BdG equations on large tight-binding models, which is based on the localized-Krylov subspace and the sparse modeling technique. 
If STM/STS or SQUID measurements can observe a vortex pinning in clean quasicrystals, this becomes evidence of the inhomogeneous superconductivity in quasicrystals.

\acknowledgments

\appendix
YN was partially supported by JSPS- KAKENHI Grant Numbers 18K11345 and 20H05278. 
The calculations were performed by the supercomputing system HPE SGI8600 at the Japan Atomic Energy Agency. 

\appendix
\section{Dual-grid method} \label{sec:dual}

\subsection{Hyper-lattices for quasicrystals}
It is known that a $D$-dimensional quasicrystal can be obtained by the projection of a particularly cut slice of the $M$-dimensional euclidian hyper-lattice onto a $D$-dimensional plane, which is expressed as 
\begin{align}
\Vec{r}_{\rm lattice} &= \sum_{\mu=0}^{M-1} K_{\mu} \Vec{d}_{\mu}, 
\end{align}
where $\vec{K} = (K_1,\cdots,K_{M})$ are the $M$ dimensional lattice points labeled with integers $K_{\mu}$. 
In this section, we use $\vec{A}$ as the $M$-dimensional vector and ${\bm a}$ as the $D$-dimensional vector. 
In the Penrose (Ammann-Beenker) lattice, the hyper-lattice is defined in $5$ (4) dimensional space. 
The $\vec{d}$ vectors in the Penrose and Ammann-Beenker lattices are respectively defined as 
\begin{align}
{\bm d}_{\mu}^{\rm Penrose} &= 
\left(\begin{array}{c}\cos \left( \frac{2 \pi \mu}{5} \right) \\
\sin \left( \frac{2 \pi \mu}{5} \right)
\end{array}\right), \\
{\bm d}_{\mu}^{\rm AB} &= 
\left(\begin{array}{c}\cos \left( \frac{\pi \mu}{4} \right) \\
\sin \left( \frac{\pi \mu}{4} \right)
\end{array}\right).
\end{align}
We consider a $M$ dimensional vector $\vec{R}$ on two dimensional plane in $M$ dimensional space defined as 
\begin{align}
\vec{R} &= x \vec{D}_1 + y \vec{D}_2 + \vec{\gamma}, \label{eq:2dplane}
\end{align}
where $\Vec{r}^{\rm T} = (x,y)$ is a coordinate on the two dimensional plane and $\vec{\gamma}^{\rm T} = (\gamma_0,\gamma_1,\gamma_2,\gamma_3,\gamma_4)$ ($\sum_{\mu} \gamma_\mu = 0$) consists of $M$ real numbers representing the shift in $M$ dimensions. 
Here, the $M$-dimensional vector $\vec{D}_i$ is defined as 
\begin{align}
\vec{D}_m^{\rm T} &= ([\Vec{d}_0]_m,\cdots,[\Vec{d}_{M-1}]_m).
\end{align}
For example, in the Penrose tiling, we have 
\begin{align}
    \vec{D}_1^{\rm T} &= [1,\cos (\theta),\cos (2\theta),\cos (3\theta),\cos (4\theta)], \\
    \vec{D}_2^{\rm T} &= [0,\sin (\theta),\sin (2\theta),\sin (3\theta),\sin (4\theta)], 
\end{align}
with $\theta \equiv 2 \pi/5$. 
If we can find the hyper-lattice point $\vec{K}$ close to the two-dimensional plane defined in Eq.~(\ref{eq:2dplane}), 
the corresponding two-dimensional real space point $\Vec{r}_{\rm lattice}$ becomes a vertex of a tiling. 

\subsection{Dual grids}
To find the hyper-lattice point $\vec{K}$ close to the two-dimensional plane, we consider the points where the crosssections of two grid lines $\alpha$ and $\beta$ are on the two dimensional plane in $M$ dimensional space, expressed as 
\begin{align}
K_{\alpha} &= \Vec{d}_{\alpha}^T \Vec{r} + \gamma_{\alpha},\\
K_{\beta} &= \Vec{d}_{\beta}^T \Vec{r} + \gamma_{\beta},
\end{align}
where $K_{\alpha}$ and $K_{\beta}$ are two integers. 
With the use of the vector $\Vec{r}$, the hyper-lattice point $\vec{K}$ can be obtained by 
\begin{align}
K_{\alpha} &= \lceil \Vec{d}_{\alpha}^T \Vec{r} + \gamma_{\alpha} \rceil.
\end{align}
We can easily generate different patterns of Penrose or Ammann-Beenker lattices with different $\vec{\gamma}$, as shown in Figs.~\ref{fig:pen}-\ref{fig:ab}. 
In other words, one can reproduce same quasicrystal structure with the use of same $\vec{\gamma}$.

\section{Details of numerical approach for large-scale superconductors} \label{sec:rscg}
\subsection{Local density of states and mean-fields}
Without diagonalizing the BdG Hamiltonian directly, we can calculate physical observable and mean-fields with the use of the one-particle Green's function defined as 
\begin{align}
    \hat{G}(\tau) &=- \langle T_{\tau} {\bm \psi}(\tau) {\bm \psi}(0)^{\dagger} \rangle,
\end{align}
where $\tau$ is imaginary time.
Here, a $2N$ component creation operator in the Nambu space ${\bm \psi}^{\dagger}$ is defined as  ${\bm \psi}^{\dagger} \equiv (c_{1 \uparrow}^{\dagger},\cdots,c_{N \uparrow}^{\dagger}, c_{1 \downarrow},\cdots, c_{N \downarrow})$ for a spin-singlet single-band superconductivity with $N$ lattice sites.
The one-particle Green's function in complex energy plane $z$ is calculated as 
\begin{align}
    \hat{G}(z) = (z \hat{I} - \hat{H})^{-1}.
\end{align}
For example, the LDOS with a quantum index $i$ ({\it e.g.} a site index or spin-index etc.) and the mean-field $\langle c_{i\downarrow} c_{i \uparrow}\rangle$ are respectively expressed as 
\begin{align}
N(\omega,i) &= -\frac{1}{2\pi i}  \Vec{e}(i)^T \hat{d}(\omega)\Vec{e}(i) ,\\
\langle c_{i\downarrow} c_{i \uparrow}\rangle &= \frac{1}{2\pi}  \int_{-\infty}^{\infty} d\omega \Vec{e}(i)^T \hat{d}(\omega) \Vec{h}(i),
\end{align}
where the difference of the retarded and advanced Green's function matrices $\hat{d}(\omega)$ is determined as $\hat{d}(\omega) = \hat{G}^{\rm R}(\omega) - \hat{G}^{\rm A}(\omega)$. 
Here, we introduce the following $2N$-component unit-vectors $\Vec{e}(i)$ and $\Vec{h}(j)$ $(1 \leq i \leq N)$, which are, respectively, defined as 
\begin{align}
[\Vec{e}(i)]_{\gamma} = \delta_{i,\gamma}, \: \: \: [\Vec{h}(i)]_{\gamma} = \delta_{i+N,\gamma}.
\end{align}
The mean-fields are also expressed with the Matsubara Green's function as\cite{NagaiRSCG} 
\begin{align}
\langle c_{i\downarrow} c_{i \uparrow}\rangle &= T \sum_{n=-\infty}^{\infty} \Vec{e}(i)^T \hat{G}(i \omega_n) \Vec{h}(i).
\end{align}
The $2N \times 2N$ matrix $\hat{G}(i \omega_n)$ is the Green's function with the Fermion Matsubara frequency $\omega_n \equiv \pi T (2n+1) $ defined as 
\begin{align}
\hat{G}(i \omega_n) &\equiv [i \omega_n \hat{I} - \hat{H}]^{-1}.
\end{align}
By solving the linear equations defined as 
\begin{align}
(i \omega_n \hat{I} - \hat{H}) \Vec{x}(i,\omega_n) &= \Vec{h}(i), \label{eq:linearAP}
\end{align}
the superconducting mean-fields are expressed as 
\begin{align}
\langle c_{i\downarrow} c_{i \uparrow}\rangle &= T \sum_{n=-\infty}^{\infty} \Vec{e}(i)^T \Vec{x}(i,\omega_n). \label{eq:xap}
\end{align}
By solving Eqs.~(\ref{eq:delta}) and (\ref{eq:xap}) self-consistently, we obtain the superconducting ground states of quasicrystals. 
Usually, one introduces a cutoff Matsubara frequency $\omega_{n_{\rm cut}}$ to approximate the above summation, where the number of the Matsubara frequencies becomes finite. 
We have introduced the reduced-shifted conjugate-gradient (RSCG) method\cite{NagaiRSCG}, which solves Eq.~(\ref{eq:xap}) with different frequencies simultaneously. 
We should note that the computational complexity based on this Matsubara formalism increases with decreasing temperature since the number of the Matsubara frequencies increases with fixing a cutoff $\omega_{n_{\rm cut}}$. 
The complexity for calculating a mean-field at site $i$ is estimated as ${\cal O}(m n_{\rm cut}) + {\cal O}(2N m)$, where the first term originates from the Matsubara summation and the second term originates from  a sparse-matrix vector operation. 
Here, $m$ is the number of the iteration steps for the RSCG. 
The total complexity for self-consistent calculation is ${\cal O}(m N n_{\rm cut}) + {\cal O}(2N^2 m)$, since we have to calculate the mean-fields everywhere.

\subsection{Sparse modeling approach: Intermediate representation for Green's functions}\label{sec:spm}
There is another method so-called sparse modeling approach (SpM)\cite{Chikano,NagaiJPSJ, Otsuki,Li,Wang}, to make the infinity Matsubara summation in Eq.~(\ref{eq:x}) computable. 
In the SpM, the intermediate representation (IR) basis is introduced to express information of Green's function. 
The IR basis originates from the Lehmann representation of the single-particle Green's function 
\begin{align}
G_{AB}(\tau) &= - \int_{-\omega_{\rm max}}^{\omega_{\rm max}} d\omega K(\tau,\omega)\rho_{AB}(\omega), \label{eq:irg}
\end{align}
where $G_{AB}(\tau)$ is defined as 
\begin{align}
    G_{AB}(\tau) = - \langle T_{\tau} A(\tau) B(0) \rangle.
\end{align}
The operator $A$ and $B$ should be one of creation or annihilation operators. 
The kernel $K(\tau,\omega)$ is defined as 
\begin{align}
K(\tau,\omega) &\equiv \frac{e^{- \tau \omega}}{1 + e^{-\beta \omega}}, 
\end{align}
for $\tau \in [0,\beta]$. 
The spectrum $\rho_{AB}(\omega)$ is defined as 
\begin{align}
    \rho_{AB}(\omega) = - \frac{1}{2\pi i} \lim_{\eta \rightarrow 0+} ( G_{AB}(\omega + i\eta) -  G_{AB}(\omega - i\eta)). 
\end{align}
Here, the spectrum  is bounded in the interval $[-\omega_{\rm max},\omega_{\rm max}]$ ($\omega_{\rm max}$ is a cutoff frequency). 
The IR basis functions are defined through the singular value decomposition (SVD) expressed as 
\begin{align}
 K(\tau,\omega) = \sum_{l}^{\infty} S_{l} U_l(\tau)V_l(\omega),
\end{align}
where the singular values $S_l$ decays exponentially with increasing $l$. 

The matrix element of the Green's function $G_{ij}$ can be expanded into a compact representation in terms of $N_{\rm IR}$ basis functions, such that in imaginary time and Matsubara frequencies, 
\begin{align}
G_{ij}(\tau) &= \sum_{l=0}^{N_{\rm IR}-1} G_{l,ij} U_l(\tau), \\
G_{ij}(i \omega_n) &= \sum_{l=0}^{N_{\rm IR}-1} G_{l,ij} U_l(i \omega_n), \\
U_l(i \omega_n) &= \int_{0}^{\beta} d\tau U_l(\tau) e^{i \omega_n \tau},
\end{align}
where $G_{l,ij}$ are expansion coefficients, $U_l(\tau)$ is the IR basis\cite{Chikano}. 
The superconducting mean-fields are expressed as 
\begin{align}
\langle c_{i\downarrow} c_{i \uparrow}\rangle &= \Vec{e}(i)^T \hat{G}(\tau = \beta) \Vec{h}(i) =
\sum_{l=0}^{N_{\rm IR}-1}  U_l(\beta) \Vec{e}(i)^T \hat{G}_l \Vec{h}(i).
\end{align}

According to Ref.~\onlinecite{Li}, if the sampling points are chosen in the distribution of the roots of the IR basis functions, 
there is a useful transformation given as 
\begin{align}
G_l &= \sum_{k=0}^{N_{\rm IR}-1} [\Vec{U}^{-1}]_{lk} G(i \omega_k),
\end{align}
where $\Vec{U}$ is a $N_{\rm IR} \times N_{\rm IR}$ matrix expressed as 
\begin{align}
[\Vec{U}]_{lk} &= U_{l}(i \omega_k).
\end{align}
The sampling points $\omega_k$ are obtained by the open-source software irbasis\cite{Chikano}. 
Finally, the superconducting mean-fields are given as 
\begin{align}
\langle c_{i\downarrow} c_{i \uparrow}\rangle &=\sum_{l=0}^{N_{\rm IR}-1} U_l(\beta) \sum_{k=0}^{N_{\rm IR}-1}  [\Vec{U}^{-1}]_{lk} \Vec{e}(i)^T \Vec{x}(i,\omega_k).
\end{align}
Although the size of $N_{\rm IR}$ depends on a cutoff parameter in the IR basis, $N_{\rm IR}$ is small around 10-100. 
With the use of the RSCG method and IR basis, we can calculate Eq.~(\ref{eq:x}) with high accuracy. 
The computational complexity for the Matsubara summations in the RSCG method ${\cal O}(m n_{\rm cut})$ is replaced with ${\cal O}(m N_{\rm IR})$.

%\bibliography{ref}

\end{document}